\documentstyle[11pt]{article}

\font\tenrm=cmr10
\font\tenit=cmti10
\font\elevenbf=cmbx10 scaled\magstep 1
\font\elevenrm=cmr10 scaled\magstep 1
\font\elevenit=cmti10 scaled\magstep 1

\renewenvironment{thebibliography}[1]
 { \elevenrm
   \begin{list}{\arabic{enumi}.}
    {\usecounter{enumi} \setlength{\parsep}{0pt}
     \setlength{\itemsep}{3pt} \settowidth{\labelwidth}{#1.}
     \sloppy
    }}{\end{list}}

\begin{document}
\begin{flushright}
FSU--HEP--930727\\
July 1993
\end{flushright}
\begin{center}{
\vglue 0.6cm
{\elevenbf RAPIDITY CORRELATIONS IN $W\gamma$ PRODUCTION \\[2.mm]
               AT THE TEVATRON
\footnote{To appear in the Proceedings of the Workshop {\it ``Physics at
Current Accelerators and the Supercollider''}, Argonne National
Laboratory, June~2 --~5, 1993.} \\}
\vglue 1.0cm
{\tenrm U. BAUR \\}
\baselineskip=13pt
{\tenit Physics Department, Florida State University\\}
\baselineskip=12pt
{\tenit Tallahassee, FL 32306\\}

\vglue 0.3cm
{\tenrm S. ERREDE\\}
{\tenit Physics Department, University of Illinois, Urbana,
IL 61801 \\}
\vglue 0.3cm
{\tenrm and\\}
\vglue 0.3cm
{\tenrm G. LANDSBERG\\}
{\tenit Physics Department, SUNY at Stony Brook, NY 11794\\}
\vglue 0.8cm
{\tenrm ABSTRACT}}

\end{center}

\vglue 0.1cm
{\rightskip=3pc
 \leftskip=3pc
 \tenrm\baselineskip=12pt
 \noindent
We study the correlation of photon and charged lepton pseudorapidities,
$\eta(\gamma)$ and $\eta(\ell)$, $\ell=e,\,\mu$, in $p\bar p\rightarrow
W^\pm\gamma+X\rightarrow \ell^\pm p\llap/_T\gamma+X$ at
the Tevatron. In the Standard Model, the $\Delta\eta(\gamma,\ell)=
\eta(\gamma) - \eta(\ell)$
differential cross section is found to exhibit a pronounced dip at
$\Delta\eta(\gamma,\ell) \approx \mp 0.4$, which originates from the
radiation zero present in $q\bar q'\rightarrow W\gamma$. The sensitivity
of the $\Delta\eta(\gamma,\ell)$ distribution to higher order QCD
corrections, non-standard $WW\gamma$ couplings, and the cuts imposed is
explored. The $\Delta\eta(\gamma,\ell)$ distribution is compared with
other quantities which are sensitive to the radiation zero.
\vglue 0.6cm}
{\elevenbf\noindent 1. Introduction}
\vglue 0.2cm
\baselineskip=14pt
\elevenrm
A pronounced feature of $W\gamma$ production in hadronic collisions is
the so-called radiation zero which appears in the parton level
subprocesses which contribute to lowest order in the Standard Model (SM) of
electroweak interactions~\cite{RAZ}. For $u\bar d\to W^+\gamma$ ($d\bar u\to
W^-\gamma$) all contributing
helicity amplitudes vanish for $\cos\Theta^*=-1/3$ (+1/3), where
$\Theta^*$ is the angle between the quark and the photon in the parton
center of mass frame. In practice, however, this zero is difficult to observe.
Structure function effects transform the zero into a dip. Higher order
QCD corrections \cite{NLO,NLOTWO} and finite $W$ width effects, together with
photon radiation from the final state lepton line, tend to fill in the
dip. Finally, the twofold ambiguity in the
reconstructed parton center of mass frame which originates from the two
possible solutions for the
longitudinal momentum of the neutrino \cite{STROUGHAIR}, $p_L(\nu)$,
represents
an additional complication in the extraction of the $\cos\Theta^*$ or
the corresponding rapidity distribution, $d\sigma/dy^*(\gamma)$, which further
dilutes the effect.

The effect of higher order QCD corrections can largely be compensated by
imposing a jet veto \cite{NLOTWO}. Unwanted effects originating from
the twofold ambiguity in $p_L(\nu)$ can be
avoided by considering quantities which reflect the radiation zero, but
which do not require the reconstruction of the parton center of mass
frame. Recently, the ratio of $Z\gamma$ and $W^\pm\gamma$ cross sections as a
function of the minimum photon $p_T$ has been demonstrated
\cite{BEO} to be such a quantity. However, a sufficiently large sample
of $Z\gamma$ events is necessary to establish the increase of the cross
section ratio with $p_T^{\rm min}(\gamma)$ predicted by the SM. In
addition one has to assume that there are no non-standard contributions
to $Z\gamma$ production, {\it e.g.} from anomalous $ZZ\gamma$ or
$Z\gamma\gamma$ couplings \cite{ELB}.

Here we consider correlations between the photon
rapidity, $\eta(\gamma)$, and the rapidity $\eta(\ell)$ of the charged
lepton, $\ell=e,\,\mu$, originating from the $W$ decay as tools to
observe the radiation zero predicted by the SM for $W\gamma$ production
in hadronic collisions. We show that the double differential
distribution $d^2\sigma/d\eta(\gamma)d\eta(\ell)$ and the distribution of
the difference of rapidities, $\Delta\eta(\gamma,\ell)=\eta(\gamma) -
\eta(\ell)$, clearly display the SM radiation zero. We also study the
sensitivity of the $\Delta\eta(\gamma,\ell)$ spectrum to higher order QCD
corrections, non-standard $WW\gamma$ couplings and the cuts imposed.
\vglue 0.3cm
{\elevenbf\noindent 2. Photon Lepton Rapidity Correlations}
\vglue 0.2cm
In our analysis we shall focus entirely on the $W^+\gamma$ channel.
Results for $W^-\gamma$ production can be obtained by exchanging the
sign of the rapidities involved. The calculation of $W\gamma$ production
in the Born approximation and at ${\cal O}(\alpha_s)$ is
performed using the results of Ref.~\cite{BB} and~\cite{NLOTWO},
respectively. In the Born approximation, our calculation fully
incorporates finite $W$
width effects, together with photon bremsstrahlung from the final state
charged lepton line. The NLO QCD calculation, on the other hand, treats
the $W$ boson in the
narrow width approximation. In this approximation, diagrams in which the photon
is radiated off the final state lepton line are not necessary to maintain
electromagnetic gauge invariance. Imposing a large photon lepton
separation cut, together with a cluster transverse mass cut, these
diagrams can be ignored~\cite{NLOTWO}.

The SM radiation zero leads to a
pronounced dip in the photon rapidity distribution in the center of
mass frame, $d\sigma/dy^*(\gamma)$, at
\begin{eqnarray}
y^*(\gamma)=y_0=-{1\over 2}\,\log 2\approx -0.35.
\end{eqnarray}
For $u\bar d\rightarrow W^+\gamma$ the photon and the $W$ are back to back
in the center of mass frame. The corresponding rapidity distribution of
the $W$ in the parton center of mass frame, $d\sigma/dy^*(W)$, thus
exhibits a dip at $y^*(W) = -y_0$. In the
double differential distribution of the rapidities in the laboratory
frame, $d^2\sigma/d\eta(\gamma)dy(W)$, one then expects a ``valley'' for
rapidities satisfying the relation\footnote{Differences of rapidities are
invariant under boosts.} $\eta(\gamma)-y(W)\equiv y^*(\gamma)-y^*(W)=2y_0$.

In the SM, the dominant $W^\pm$ helicity in $W\gamma$ production is
$\lambda_W = \pm 1$~\cite{BBS}, implying that
the charged lepton will tend to be emitted in the direction of the parent
$W$, thus reflecting most of its kinematic properties. The difference
in rapidity, $\Delta y(W,\ell)=y(W)-\eta(\ell)$,
between the $W$ boson and the charged lepton originating from the $W$ decay
is rather small with an average $\Delta y(W,\ell)$ of 0.30.
The double differential distribution $d^2\sigma/d\eta(\gamma)d\eta(\ell)
$ for $p\bar p\rightarrow W^+\gamma\rightarrow\ell^+p\llap/_T\gamma$
is thus expected to display a valley for rapidities fulfilling the relation
$\Delta\eta(\gamma,\ell)=\eta(\gamma) -\eta(\ell)\approx -0.4$.
Figure~1 shows $d^2\sigma/d\eta(\gamma)d\eta(\ell^+)$ for
$p\bar p\rightarrow \ell^+p\llap/_T\gamma$ in the Born approximation,
together with the
corresponding distribution for $p\bar p\rightarrow \ell^+\ell^-\gamma$.
\begin{figure}[t]
\vskip 3in
\includegraphics{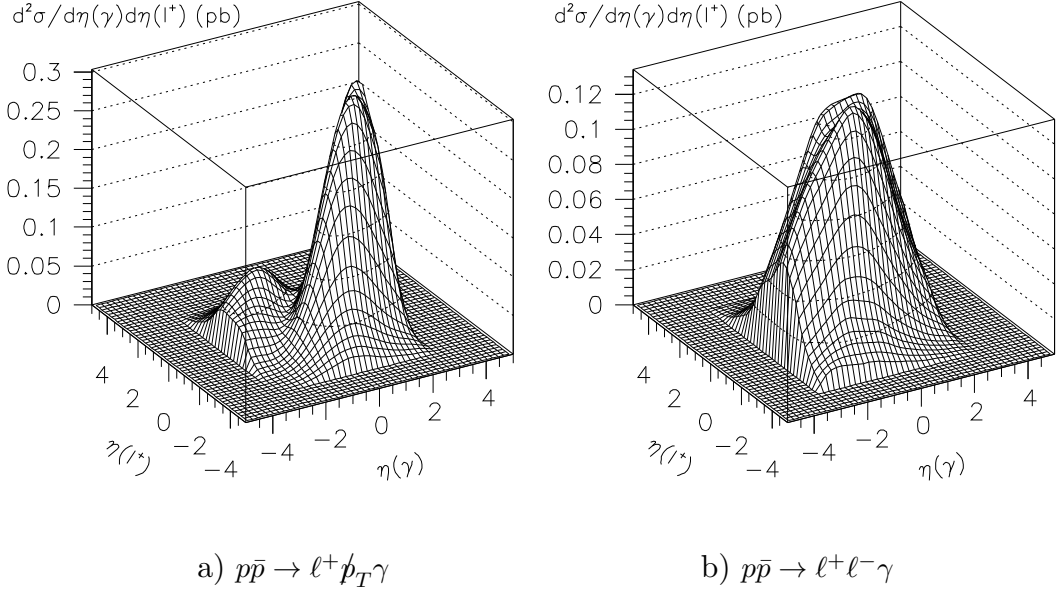}
\hglue 3.cm a) $p\bar p\rightarrow \ell^+p\llap/_T\gamma$ \hglue 4.cm b) $p\bar
p\rightarrow \ell^+\ell^-\gamma$ \\
\caption{The double differential distribution $d^2\sigma/d\eta(\gamma)
d\eta(\ell^+)$, $\ell=e,\,\mu$, for a) $p\bar p\rightarrow
\ell^+p\llap/_T\gamma$ and b) $p\bar
p\rightarrow \ell^+\ell^-\gamma$ at the Tevatron. The cuts imposed are
described in the text.}
\end{figure}
Here we have imposed a $p_T(\gamma)>5$~GeV, a $p_T(\ell)>20$~GeV and a
$p\llap/_T>20$~GeV cut, together with cuts on the pseudorapidities of
the photon and charged lepton of $|\eta(\gamma)|<3$ and $|\eta(\ell)| <
3.5$. To select a phase space region where radiative $W$ ($Z$) decays
are suppressed and $q\bar q'\rightarrow W\gamma$ ($q\bar q\rightarrow
Z\gamma$) dominates, we have required in addition a large photon lepton
separation cut of $\Delta R(\gamma,\ell)>0.7$, a cluster transverse
mass cut of
$m_T(\ell\gamma;p\llap/_T) > 90$~GeV for $W^+\gamma$ production, and
invariant mass cuts of $m(\ell^+\ell^-)>70$~GeV and
$m(\ell^+\ell^-\gamma)>100$~GeV in the $Z\gamma$ case. For the parton
distribution functions we use the MRSS0 parametrization \cite{MRS}. The double
differential cross section for $p\bar p\rightarrow\ell^+\ell^-\gamma$ is
calculated using the results of Ref.~\cite{ELB}.

Figure~1a exhibits a pronounced minimum at $\Delta\eta(\gamma,\ell)
\approx -0.4$, as expected. Furthermore, the photon and lepton
rapidities are seen to be strongly correlated, with most photons
(leptons) having positive (negative) rapidities. Since the sign of the
rapidities changes for $W^-\gamma$ production, this correlation may aid
in determining the charge of the electron in the D\O \
detector which does not have a central magnetic field. If the cluster
transverse
mass cut of $m_T(\ell\gamma;p\llap/_T)>90$~GeV is removed, radiative $W$
decays dominate and the valley disappears. In contrast to the situation
for $p\bar p\rightarrow\ell^+p\llap/_T\gamma$, there is no sign of a
valley, and no strong correlation between the rapidities in $p\bar
p\rightarrow\ell^+\ell^-\gamma$ (see Fig.~1b).

Since the valley is approximately 1.5 units in rapidity wide and occurs
essentially in the diagonal of the $\eta(\gamma),\eta(\ell)$ plane, it
is completely obscured if one integrates over the full range of either
the photon or the lepton
rapidity. Only for sufficiently strong cuts, {\it e.g} $|\eta(\gamma)|<1$ or
$|\eta(\ell)|<1$, a slight dip can be observed in
$d\sigma/d\eta(\ell)$ ($d\sigma/d\eta(\gamma)$) in the region around
$\eta(\ell)\approx 0.4$ ($\eta(\gamma)\approx -0.4$).

The double differential distribution $d^2\sigma/d\eta(\gamma)d\eta(\ell)
$ can only be mapped out if a sufficiently large number of events is
available. For a relatively small event sample the distribution of the
rapidity difference, $d\sigma/d\Delta\eta(\gamma,\ell)$, is more useful. The
$\Delta\eta(\gamma,\ell)$ distribution for the cuts summarized above is
shown in Fig.~2a (solid line). As anticipated, the distribution exhibits a
strong dip at $\Delta\eta(\gamma,\ell)\approx -0.4$.
\begin{figure}[t]
\vskip 9.5cm
\includegraphics{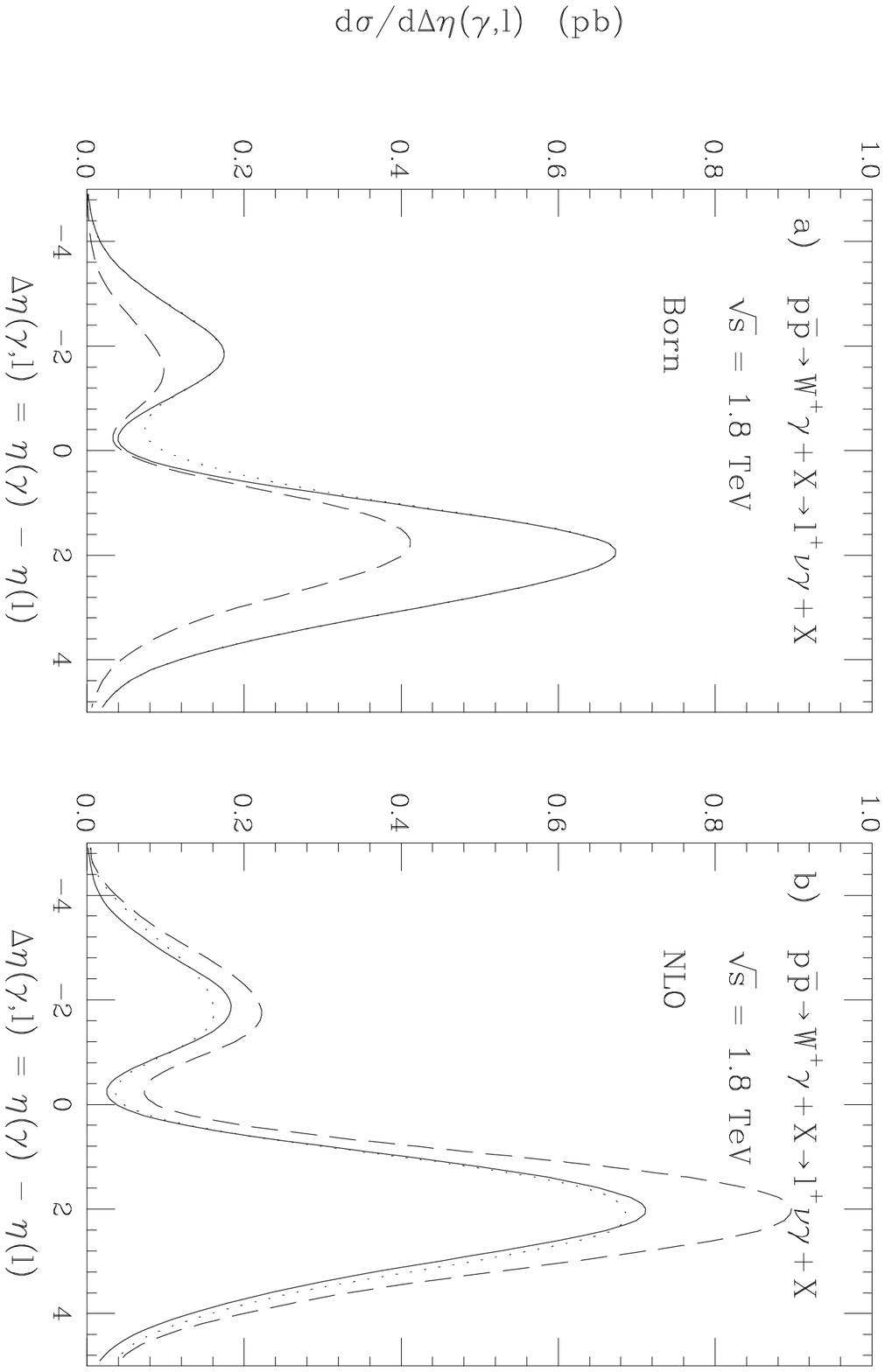}
\noindent Figure~2: a) The rapidity difference distribution,
$d\sigma/d\Delta\eta(\gamma,\ell)$, for $p\bar p\rightarrow
W^+\gamma+X\rightarrow\ell^+p\llap/_T\gamma+X$,  $\ell=e,\,\mu$, in the
Born approximation.
The solid line shows the result obtained for the cuts described in the
text. The dashed (dotted) curve displays the rapidity difference distribution
if the $p_T(\gamma)>5$~GeV ($\Delta R(\gamma,\ell)>0.7$) cut is replaced
by $p_T(\gamma)>10$~GeV ($\Delta R(\gamma,\ell)>0.3$), with all other
cuts unchanged.\\
b) The $\Delta\eta(\gamma,\ell)$ distribution, including ${\cal
O}(\alpha_s)$ QCD corrections, for the cuts described in the text. The
solid and dashed lines represent the
Born and the inclusive NLO result, respectively. The dotted line shows
the NLO result obtained for the exclusive reaction $p\bar p\rightarrow
W^+\gamma+0$~jet, $W^+\rightarrow\ell^+\nu$.
\end{figure}
In the region of the dip, most events originate from the high
$p_T(\gamma)$ region. This is illustrated by the dashed line in Fig.~2a,
which shows the $\Delta\eta(\gamma,\ell)$ distribution for $p_T(\gamma)
>10$~GeV instead of $p_T(\gamma)>5$~GeV. Around the
minimum the result almost coincides with that obtained for the smaller
photon $p_T$ cut.
Increasing the photon transverse momentum threshold, the dip becomes less
pronounced. The dotted line in Fig.~2a, finally,
displays the rapidity difference distribution with the $\Delta R(\gamma,
\ell)>0.7$ cut replaced by $\Delta R(\gamma,\ell)>0.3$. Reducing the
photon lepton isolation cut increases the contribution of the diagram
where the photon is radiated off the final state lepton line. The final
state bremsstrahlung contribution, which diverges in the collinear
limit, destroys the SM radiation zero, and thus tends to fill in the
dip.

The $\Delta\eta(\gamma,\ell)$ distribution at next-to-leading order in
QCD is shown in Fig.~2b. Besides the cuts described above, we also
require the photon to be isolated from hadrons in the NLO calculation by
imposing a cut on the total hadronic energy in a cone of size $\Delta
R=0.7$ about the direction of the photon of
\begin{eqnarray}
\sum_{\Delta R < 0.7} \, E_{\hbox{\scriptsize had}} <
0.15 \, E_{\gamma} \>,
\label{EQ:ISOL}
\end{eqnarray}
where $E_\gamma$ is the photon energy. This requirement strongly reduces
photon bremsstrahlung from final state quarks and gluons.

NLO QCD corrections are seen to partially
fill in the dip caused by the SM radiation zero (dashed line). While
${\cal O}(\alpha_s)$ QCD corrections enhance the cross section by 30 --
40\% outside the dip region, they increase the rate by approximately a
factor 2.5 at $\Delta\eta(\gamma,\ell)\approx -0.4$. This effect is
predominantly caused by the $2\rightarrow 3$ processes $qg\rightarrow
W\gamma q'$ and $\bar q'g\rightarrow W\gamma\bar q$ where no radiation
zero is present in the helicity amplitudes. Imposing a jet veto, {\it
i.e.} requiring that no jets with transverse momentum $p_T(j)>10$~GeV and
rapidity $|\eta(j)|<2.5$ are present in the event, the NLO
$\Delta\eta(\gamma,\ell)$ distribution (dotted line) is very
similar to that obtained in the Born approximation (solid line).

In $W\gamma$ production both the virtual $W$ and the
decaying onshell $W$ couple to essentially massless fermions, which
insures that effectively $\partial_\mu W^\mu=0$. This condition
together with Lorentz invariance, electromagnetic gauge invariance, and $CP$
conservation, allows two free parameters, $\kappa$ and $\lambda$,
in the $WW\gamma$ vertex. The most general vertex compatible with these
conservation laws is described by the effective Lagrangian~\cite{LAGRANGIAN}
\begin{eqnarray}
\noalign{\vskip 2pt}
{\cal L}_{WW\gamma} &=& -i \, e \,
\Biggl[ W_{\mu\nu}^{\dagger} W^{\mu} A^{\nu}
              -W_{\mu}^{\dagger} A_{\nu} W^{\mu\nu}
+ \kappa W_{\mu}^{\dagger} W_{\nu} F^{\mu\nu}
+ {\lambda \over M_W^2} W_{\lambda \mu}^{\dagger} W^{\mu}_{\nu} F^{\nu\lambda}
\Biggr] \>,
\label{EQ:LAGRANGE}
\end{eqnarray}
where $A^{\mu}$ and $W^{\mu}$ are the photon and $W^-$ fields, respectively,
$W_{\mu\nu} = \partial_{\mu}W_{\nu} - \partial_{\nu}W_{\mu}$, and
$F_{\mu\nu} = \partial_{\mu}A_{\nu} - \partial_{\nu}A_{\mu}$.
The variables $\kappa$ and $\lambda$ are related to the magnetic dipole
moment, $\mu_W^{}$, and the electric quadrupole moment, $Q_W^{}$, of
the $W$-boson:
\begin{eqnarray}
\mu_{W}^{} = {e \over 2 M_W^{} }\, (1 + \kappa + \lambda) \>, \hskip
1.cm
Q_{W}^{} = -{e \over M_W^2 } \, (\kappa - \lambda) \>.
\end{eqnarray}
At tree level in the SM, $\kappa = 1$ and $\lambda = 0$.
The two $CP$ conserving couplings have recently been measured by the UA2
Collaboration in the process $p\bar p\rightarrow e^\pm\nu\gamma X$ at
the CERN $p\bar p$ collider~\cite{UA2}:
\newcommand{\crc}{\crcr\noalign{\vskip -3pt}}
\begin{eqnarray}
\noalign{\vskip 5pt}
\kappa=1\matrix{+2.6\crc -2.2}~~({\rm for}~\lambda=0) \>, \hskip 1.cm
\lambda=0\matrix{+1.7\crc -1.8}~~({\rm for}~\kappa=1) \>,
\end{eqnarray}
at the 68.3\% confidence level (CL). Although bounds on these couplings
can also be extracted from
low energy data and high precision measurements at the $Z$ pole, there are
ambiguities and model dependencies in the results \cite{LOW}.
No rigorous bounds on $WW\gamma$ couplings can be
obtained from LEP~I data if correlations between different
contributions to the anomalous couplings are fully taken into account.

Tree level unitarity uniquely restricts the $WW\gamma$ couplings to their
SM gauge theory values at asymptotically high energies~\cite{CORNWALL}.
This implies that any deviation of $\kappa$ or $\lambda$ from the SM
expectation has to be described by a form factor
$a(M_{W\gamma}^2, p_W^2, p_\gamma^2)$, $a=(\Delta\kappa=\kappa-1),\,
\lambda$ which
vanishes at high energies. Consequently, the anomalous couplings are introduced
via form factors
\begin{eqnarray}
\noalign{\vskip 2pt}
a(M_{W\gamma}^2, p_W^2 = M_W^2, p_\gamma^2 = 0) \> &=& \>
{a_0 \over (1 + M_{W\gamma}^2/\Lambda^2)^n } \>,
\label{EQ:KAPPAFORM}
\end{eqnarray}
where $a_0=\Delta \kappa_0,\,\lambda_0$ are the form factor values at
low energies and
$\Lambda$ represents the scale at which new physics becomes important in the
weak boson sector, {\it e.g.} due to a composite structure of the $W$-boson.
For the numerical results presented here, we use a dipole
form factor ($n=2$) with a scale $\Lambda = 1$~TeV.

The sensitivity of the rapidity difference distribution in the Born
approximation to non-standard $WW\gamma$ couplings is explored in
Fig.~3. The solid line shows the SM result, whereas the dashed and
dotted curves give the prediction for the current UA2 68\% CL limits of
$\Delta\kappa_0=2.6$ and $\lambda_0=1.7$, respectively.
\begin{figure}[t]
\vskip 9.5cm
\includegraphics{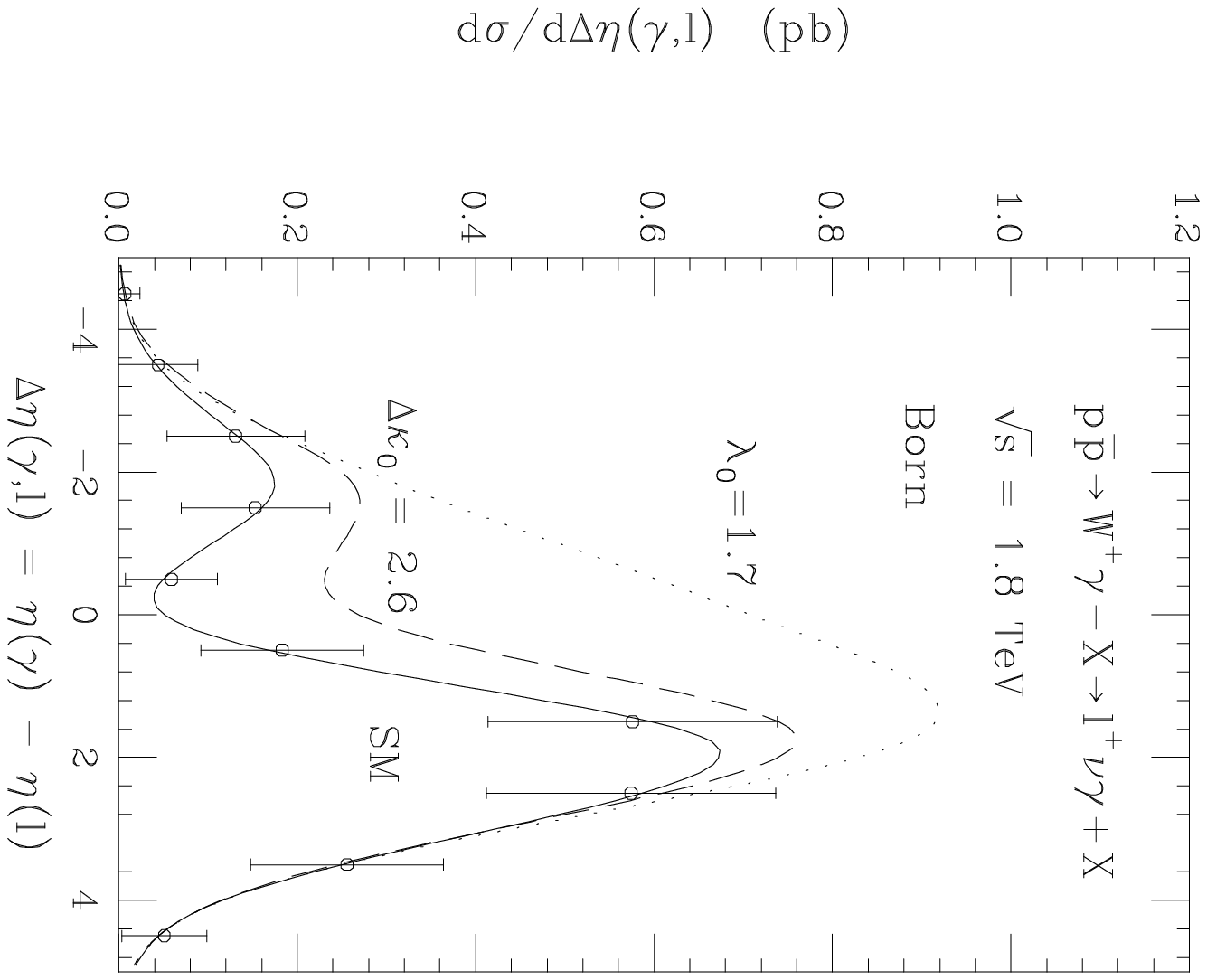}
\noindent Figure~3: The rapidity difference distribution,
$d\sigma/d\Delta\eta(\gamma,\ell)$, for $p\bar p\rightarrow
W^+\gamma+X\rightarrow\ell^+p\llap/_T\gamma+X$,  $\ell=e,\,\mu$, at
the Tevatron in the Born
approximation for anomalous $WW\gamma$ couplings. The curves are for
the SM (solid), $\Delta\kappa_0=2.6$
(dashed), and $\lambda_0=1.7$ (dotted). Only one coupling is varied at a
time. A dipole form factor with scale
$\Lambda=1$~TeV is assumed for non-standard $WW\gamma$ couplings. The
cuts imposed are described in the text. The error bars indicate the
expected statistical errors for an integrated luminosity of
22~pb$^{-1}$.
\end{figure}
In presence of any anomalous contribution to the $WW\gamma$ vertex the
radiation zero is eliminated and the dip in $d\sigma/d\Delta\eta(\gamma,
\ell)$ is filled in at least partially. Most of the excess cross section
for non-standard couplings originates in the high $p_T(\gamma)$ region
\cite{BB}, where events tend to be central in rapidity. Deviations from
the SM $\Delta\eta(\gamma,\ell)$ distribution, therefore, mostly occur
for small rapidity differences. In Fig.~3 we have also included the
statistical errors expected in the SM case for an integrated luminosity
of $\int\!{\cal L} dt=22$~pb$^{-1}$. This demonstrates that the rapidity
difference distribution is sensitive to anomalous $WW\gamma$ couplings
already with the current CDF and D\O \ data samples, in particular
to $\lambda$. However, we do not expect $d\sigma/d\Delta\eta(\gamma,
\ell)$ to be more sensitive to anomalous couplings than the photon
transverse momentum distribution.
\vglue 0.3cm
{\elevenbf\noindent 3. Conclusions}
\vglue 0.2cm
We have considered photon -- lepton rapidity correlations as a tool to
study the radiation zero predicted by the SM for $W\gamma$ production
in hadronic collisions. In the SM, the dominant $W^\pm$ helicity in
$W\gamma$ production is $\lambda_W=\pm 1$. Combined with the $V-A$ coupling of
the charged lepton to the $W$, this implies that the lepton tends to be
emitted in the direction of the parent $W$, thus reflecting most of its
kinematic properties. As a result we found that the SM radiation zero
leads to a pronounced valley in the double differential distribution,
$d^2\sigma/d\eta(\gamma)d\eta(\ell)$, for $W^\pm\gamma$ production and
rapidities fulfilling the relation $\Delta\eta(\gamma,\ell)=
\eta(\gamma)-\eta(\ell)\approx \mp 0.4$.

Equivalently to the double differential distribution, the rapidity
difference distribution, $d\sigma/d\Delta\eta(\gamma,\ell)$, can be
studied. Here the radiation zero is signaled by a dip located at
$\Delta\eta(\gamma,\ell)\approx\mp 0.4$. The details of the rapidity
difference distribution are sensitive to the cuts imposed;
increasing the photon $p_T$ threshold and reducing the photon lepton
isolation cut tends to fill in the dip. A similar trend is observed for
non-standard $WW\gamma$ couplings, and if
NLO QCD corrections are taken into account. However, if the a jet veto is
imposed, {\it i.e.} the exclusive $W\gamma+0$~jet channel is considered,
the NLO rapidity difference distribution is very similar to that
obtained in the Born approximation.

Compared to $d\sigma/dy^*(\gamma)$, the rapidity difference distribution
has the advantage of being independent of the twofold ambiguity in the
reconstruction of the parton center of mass frame, which partially
obscures the radiation zero in the $y^*(\gamma)$ distribution. In
contrast to the $Z\gamma$ to $W^\pm\gamma$ cross section ratio which
also reflects the radiation zero, the rapidity difference distribution
does not depend on the $Z\gamma$ cross section, and the validity of the
SM for $p\bar p\rightarrow Z\gamma$.

While the NLO QCD corrections to $W\gamma+0$~jet production are small at
the Tevatron, this is {\it not} the case for realistic jet definitions
at supercollider energies, due to the very much increased $qg$
luminosity. For a jet defining $p_T$ threshold of 30~GeV or larger, the
dip is completely filled in at the SSC. Present studies \cite{SDC} suggest
that it will
be difficult to reconstruct jets at the SSC with a transverse momentum
less than about 30~GeV. Given a sufficiently large integrated luminosity,
experiments at the Tevatron studying photon -- lepton rapidity correlations
thus offer a {\it unique} chance to search for the SM
radiation zero in hadronic $W\gamma$ production.
\vglue 0.3cm
{\elevenbf\noindent 4. Acknowledgements}
\vglue 0.2cm
We would like to thank P.~Grannis and D.~Zeppenfeld for stimulating
discussions. This research was supported in part by the
U.~S.~Department of Energy under Grant No.~DE-FG02-91ER40677 and
Contract No.~DE-FG05-87ER40319.
\vglue 0.3cm
{\elevenbf\noindent 5. References}
\vglue 0.2cm

\end{document}